\documentclass[letterpaper,journal,9pt]{IEEEtran}

\usepackage{amsmath,amsfonts}
\usepackage{array}
\usepackage{booktabs}
\usepackage{tabularx}
\usepackage{ragged2e}
\usepackage[caption=false,font=normalsize,labelfont=sf,textfont=sf]{subfig}
\usepackage{textcomp}
\usepackage{stfloats}
\usepackage{url}
\usepackage{graphicx}
\usepackage{cite}
\usepackage{microtype}
\usepackage{balance}

\newcolumntype{Y}{>{\RaggedRight\arraybackslash}X}

\begin{document}

\title{Assurance-Scoped Reliability for Agentic Networks: \\
Capturing the State That Matters}

\author{Bilgehan Erman, Andrea Francini, and Nikos Papadis}%

\markboth{Work submitted to IEEE Communications Magazine for possible publication}
{Erman \MakeLowercase{\textit{et al.}}: Assurance-Scoped Reliability for Agentic Networks}

\maketitle

\begin{abstract}
Agentic networks transform accepted intents into operational services through autonomous reasoning, adaptive planning, tool use, and cross-domain coordination, but these capabilities introduce failure modes that conventional reliability measures do not fully capture. An accepted intent may still be carried out incorrectly, for example because the system acts on stale information, repeats an external action, applies only part of a change, or enters a fallback mode that quietly relaxes policy enforcement. Such failures can leave a service running and apparently healthy while its behavior is unsafe or unaccountable, with too little evidence to detect, explain, or recover from them. This article proposes Reliability Assurance Intelligence (RAI), a general assurance architecture for such systems. From the service description, RAI derives a per-service reliability profile that states what must be checked, recorded, recovered, and audited. At runtime, generic functions use the service profile to retain the durable state needed for recovery and accountability in a context capsule adapted to service conditions. Using an agentic lifecycle manager for deterministic network services as a running example, we design the RAI architecture and propose a methodology for validating its reliability assurances.
\end{abstract}

\section{Introduction}

\IEEEPARstart{A}{dvanced} agentic systems move beyond scripted operation toward goal-directed automation. Unlike predefined control loops with fixed corrective actions, they can interpret goals, choose among possible courses of action, and revise plans during execution. This changes the reliability problem, because correct operation now depends not only on outcomes, but also on whether autonomous decisions remain authorized, recoverable, and accountable.

In networked systems, two converging trends are bringing this form of autonomy into practice. Intent-based operation allows an operator to state a goal, such as connecting two sites with bounded latency, while leaving realization to the management system. This shift is already reflected in standards on autonomous network architecture and intent-based service assurance~\cite{itu_y3061,rfc9417}. At the same time, agentic AI is entering the management system. It can interpret intents, select tools, coordinate across domains, and revise plans while retaining memory across long-running workflows. The assurance problem considered here arises whenever a management system combines intent interpretation, adaptive planning, external tool calls, and long-lived state. Agentic AI makes this combination systematic.

The resulting problem extends the classical dependability vocabulary of faults, failures, availability, safety, and integrity~\cite{avizienis_dependable_2004} with assurance questions about intent traceability, authority continuity, accountable commit boundaries, and the evidence that must survive for recovery and audit.
For example, a locally reasonable retry can be globally harmful; a plan can meet its goal while breaking a time or cost budget; a model-generated remediation can be plausible yet based on stale observations; and a degraded mode can keep a service alive while silently bypassing a policy check or leaving too little evidence to explain why it was authorized.

A reliable agentic management system must therefore demonstrate more than the availability of its components. It must preserve a trace from the accepted intent through planning and execution, and show that valid authority was retained, that externally visible actions crossed an accountable commit boundary, and that sufficient evidence survives for recovery or audit. We call this collection of demonstrations \emph{assurance}. By \emph{durable state}, we mean records that persist beyond the agent's volatile execution context, with retention and replication set by the managed service's assurance obligations.
We focus on operational, non-malicious failures, such as crashes, stale observations, duplicated tool calls, partial commits, policy brownouts, and missing recovery evidence. Byzantine faults, deliberate evidence fabrication, storage compromise, and attacks on the assurance layer remain out of scope, as do cryptographic integrity and non-repudiation mechanisms.

\begin{figure}[!t]
\centering
\includegraphics[width=0.92\columnwidth]{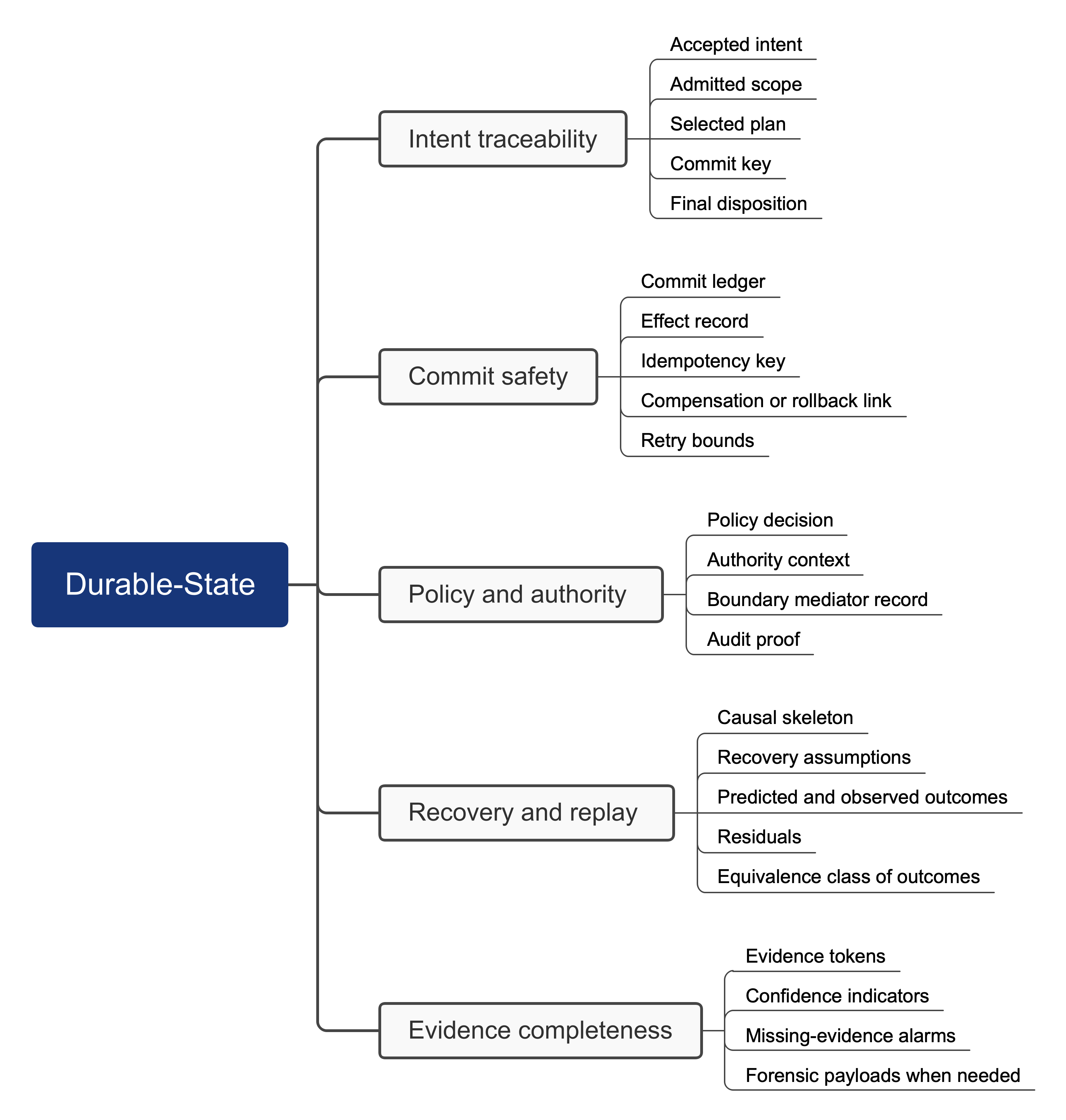}
\caption{Five families of assurance-relevant durable state.}
\label{fig:durable-state}
\end{figure}

Durable state is the practical bottleneck through which we explain the broader framework. It is where decisions about intents, commits, policy, recovery, and audit must be preserved strongly enough to support assurance obligations. Fig.~\ref{fig:durable-state} groups these obligations into five families. The design task is to identify which records carry assurance value and how that choice should adapt to risk without exhausting storage, bandwidth, and operator attention.

We propose Reliability Assurance Intelligence (RAI) as a reusable architecture for identifying and retaining the records needed for assurance. RAI combines design-time artifacts and runtime functions. The Agentic Service Calculus (ASC) provides a common behavioral vocabulary together with rules for composing behaviors; the Agentic Service Description (ASD) specifies what a service may do; and the Agentic Service Reliability Profile (ASRP) states what must be assured. At runtime, a context capsule retains the profile-selected records, an observer checks emitted evidence, and a bounded capsule controller adjusts how much evidence the capsule records as conditions change. 

This article makes three contributions. First, it introduces a general chain of design-time artifacts and runtime functions, composed of ASC, ASD, ASRP, observer, and capsule controller, for reliability reasoning across heterogeneous agentic services. 
Second, it separates what a service may do from the service-specific assurance obligations it must satisfy. 
Third, it illustrates how reusable RAI runtime functions can apply those obligations to assurance-scoped durable-state management. 
The first two contributions are general, while durable-state control is the primary application through which we show how they work in detail (it is not part of the definition of RAI itself). 
In keeping with this scope, we contribute the design of the RAI architecture and a concrete validation methodology.

\section{Running Example: An Agentic Lifecycle Manager}

Throughout the article, we illustrate RAI through a running example based on a lifecycle manager for deterministic networking services. A deterministic networking framework provides bounded latency and zero loss to packet flows that require such guarantees. It maintains these per-flow guarantees through a combination of service management and data-plane traffic management functions~\cite{francini_daas_2024}.

In our running example, the service management system is agentic; we call it the \emph{agentic lifecycle manager}. The lifecycle manager is not part of RAI, but the system under assurance. RAI represents its possible behavior in an ASD, derives its service-specific assurance obligations in an ASRP, and checks its emitted evidence at runtime through the observer and capsule controller. Where later figures (Figs.~\ref{fig:rai-pipeline} and~\ref{fig:runtime-loop}) use generic labels such as \emph{service} or \emph{service agents}, they refer to this system under assurance. The deterministic flows that the manager admits are the \emph{managed deterministic services}.

The lifecycle manager accepts an intent such as ``provide deterministic connectivity between two endpoints with a specified delay bound for a specified duration.'' It observes topology, resources, and policy state, then plans a feasible path and allocation, possibly using a learned model. If a feasible plan exists, it admits the request, reserves resources, commits the accountable binding, programs the network elements, and monitors the running service. It also handles renegotiation, degradation, compensation, and teardown, and may coordinate with peer managers as conditions change. Long-lived memory is essential because a renegotiation hours later must still be linked to the originating intent.

The manager is the source of truth about which managed services exist, which resources are committed, and which obligations remain open.Its durable state is therefore richer than the per-flow state of any single service instance, spanning the context in which decisions were made, the lifecycle by which services and resources were committed, and the outcome each service eventually reached. It records the governing context through policy versions, model or planner versions, and the freshness of the observations on which decisions relied. It also records the service and resource lifecycle, including tentative and confirmed reservations, commit keys, and confirmation that the network actually applied committed changes. Finally, it records service outcomes, including degraded-mode transitions and whether each service ended in normal completion, teardown, or failure with compensation. This richer durable state raises a practical question: \emph{which parts of it must survive, and how should retention change as assurance risk changes?}

\section{RAI: From Logs to Assurance}
Operational observability uses metrics, traces, and event streams to answer ``what happened?'' Assuring an agentic lifecycle manager requires broader, accountability-oriented questions about a given service episode: whether the right authority existed, whether a commit cleanly separated tentative planning from accountable execution, and whether degraded operation still preserved the required evidence. Just as important, agentic failures often surface not as explicit errors but as absences of evidence, such as a missing commit for an external effect or no proof that an intent was ever admissible.

The interpretation of missing evidence therefore changes. In conventional operations, a missing log line is often treated as an observability gap. In an agentic system, it can instead be an assurance gap, indicating that the system lacks the records needed to justify, recover, or safely complete its actions. This appears in several common situations. If the manager reserves resources in a peer domain and later revises the plan, the operator must know whether the old reservation was released or compensated. If a policy service becomes unavailable, the operator must know whether a still-valid cached policy authorized a restricted mode. If a tool call is retried, the operator needs an \emph{idempotency key}, an identifier reused across retries so the receiver can recognize a duplicate and apply the effect only once. In each case, the issue is not simply that some data are missing, but that the missing record can prevent accountable recovery.

Table~\ref{tab:assurance-state} summarizes records that may carry assurance value. It is not a permanent retention list, but a catalog from which a service-specific assurance policy can later select.

\begin{table*}[!t]
\caption{Assurance-Relevant State in an Agentic Management System}
\label{tab:assurance-state}
\centering
\footnotesize
\renewcommand{\arraystretch}{1.04}
\begin{tabularx}{\textwidth}{p{0.26\textwidth}YY}
\toprule
Reliability challenge & Why conventional observability is incomplete & Records with assurance value \\
\midrule
Intent lifecycle and plan variability & The same intent can lead to different plans as context, model output, or coordination changes. & Accepted intent, admitted scope, selected plan and version, commit key, final disposition. \\
Partial observability and stochastic decisions & No component sees the full state, and probabilistic choices make diagnosis ambiguous. & Observations used, freshness, model or planner version, confidence indicator, missing-evidence alarm. \\
Retries and external effects & A retry can duplicate cost or physical impact unless effects are bounded and keyed. & Commit ledger, effect record, idempotency key, confirmation, compensation or rollback link. \\
Cross-domain policy and authority & Meaning, control, and authority can change at administrative boundaries. & Boundary record, policy decision and version, authority context, mediated evidence. \\
Degraded operation under stress & Keeping a service alive can be unsafe if degradation bypasses safety or human authority. & Degradation trigger, restricted contract, cached-policy validity, escalation and recovery conditions. \\
Runtime drift and emergent interaction & Risk shifts with load, retries, latency, and multi-agent behavior. & Drift signals, capsule-mode history, residuals, bounded forensic payload. \\
\bottomrule
\end{tabularx}
\end{table*}

\subsection{What RAI Adds}
The records in Table~\ref{tab:assurance-state} do not require an entirely new capture stack. Existing techniques already provide several mechanisms: distributed tracing reconstructs causal request paths~\cite{sigelman_dapper_2010}, provenance models preserve derivation and responsibility~\cite{w3c_prov_dm_2013}, and runtime verification
checks executions against specified properties~\cite{leucker_runtime_verification_2009}, while logging, policy-as-code, and audit records cover event capture, rule expression, and evidence retention. 
RAI builds on these rather than replacing them, specifying which to apply and when.

RAI's innovation is the normalization-based artifact chain that separates service-specific behavior and obligations from generic assurance machinery. Heterogeneous services are written in the common ASC vocabulary and compiled into service-specific ASRPs; the same generic type checks, durable-state selection, observer logic, and capsule control then apply across services, with each ASRP supplying the obligations and parameters that specialize them.

The ASD records possible behavior; the ASRP identifies which parts create recovery, authority, commitment, and evidence obligations; and the runtime layer maps those obligations to required records and bounded recording modes. RAI thus addresses not only what happened or whether a fixed property held, but also which evidence must survive for a service, why, and how the evidence policy may adapt without breaking the assurance contract. 
The innovation lies in the normalization-based artifact chain, which supports derivation and reuse across heterogeneous services.

\subsection{The RAI Artifact Chain}
\subsubsection{Assurance Scope}
Four requirements define an assurance-scoped reliability profile.

\emph{Intent lifecycle continuity.} Once an intent is received, the system must be able to show whether it was admitted or rejected, how it was planned, which accountable actions and effects followed, and how it closed. Internal prompts and candidate plans need not be kept, but a durable chain must link the intent to its final outcome: completion, rejection, degradation, escalation, or compensation.

\emph{Commit safety.} Agentic systems mix tentative reasoning with actions that affect resources, users, or infrastructure. A durable commit point marks the boundary where planning becomes accountable execution; it must record the authorization, the external effect, an idempotency key when retries are possible, and any compensation obligation. Without it, after a failure the system cannot reliably reconstruct the effect needed for safe recovery. Existing systems can store such records durably~\cite{balakrishnan_logact_2026}; RAI's role is to specify which must be retained.

\emph{Policy and authority continuity.} Whether an action is permitted can depend on the service, the responsible administrative domain, the operating mode, the risk level, and whether a human is supervising. The system must therefore keep a durable record of the policy in force and its decision. A reduced fallback mode may narrow scope, automation, or data freshness, but it must never silently weaken the safeguards for safety, authorization, or auditability.

\emph{Resource-bounded evidence sufficiency.} Replication, coordination, and durable storage consume resources. A service's assurance profile must separate evidence mandatory for recovery or accountability from detail useful mainly for diagnosis. The goal is to retain enough evidence for recovery and accountability within the resource budget.

\subsubsection{ASC: A Common Language for Service Description}
For an assurance profile to apply regardless of how a service is built, RAI needs a common, implementation-independent way to describe service behavior. We introduce the \emph{Agentic Service Calculus} (ASC), a small modeling language for describing and composing service behavior in an implementation-independent way. ASC uses familiar building blocks, such as intents, actions, policy guards, domains, operating modes, commits, evidence, and resource bounds. It describes each step of a service, associates steps with the evidence they should produce, and provides composition rules for combining behaviors coherently. A lightweight type system helps catch inconsistent descriptions early.

ASC borrows proven ideas from formal languages, including compact, composable descriptions with precise behavioral comparison~\cite{anderson_netkat_2014,bergstra_thread_2007,rai_algebraic_2015,hu_service_2014} and a clean separation between partial observability and uncertain action outcomes~\cite{kaelbling_pomdp_1998}. ASC extends that basis with the concepts needed for assurance: durable commit boundaries, evidence-emitting behavior, policy and authority context, explicit degraded modes, and bounded assurance work. 
This article focuses on how those concepts are used and what they let RAI check. 
The language's full grammar and semantics are outside its scope.

The following fragment uses ASC to describe the agentic lifecycle manager of our running example, following one possible sequence of steps it may take to handle an intent. Generated at onboarding, it is a design-time description of what the service may do:

\begin{center}
\begin{minipage}{0.90\columnwidth}
\scriptsize\ttfamily
?policy\_ok(intent, policy\_v);\\
admit(intent);\\
plan(plan\_id, model\_v, freshness);\\
reserve(resource, idem\_key);\\
commit(commit\_key);\\
!\{intent\_id, policy\_v, plan\_id, model\_v,\\
\hspace*{1.5em}freshness, idem\_key, commit\_key\}.
\end{minipage}
\end{center}

The fragment is one illustrative lifecycle path. 
Read step by step, at runtime the manager would check the applicable policy, admit the intent, select a versioned, freshness-qualified plan, reserve a resource under a stable retry key, cross a keyed commit boundary, and emit the evidence needed to justify and recover the decision.

The notation is deliberately small. 
A leading \(?\phi\) is a test: the path proceeds only when the condition \(\phi\) holds. 
A semicolon means one step follows another. 
The terms \texttt{admit}, \texttt{plan}, and \texttt{reserve} are service-specific actions. 
The term \texttt{commit(k)} marks a durable accountability boundary that separates tentative reasoning from actions with external consequences. 
A leading \texttt{!} emits an evidence record. 
Suffixes such as \texttt{\_v}, \texttt{\_id}, and \texttt{\_key} denote versions, identifiers, and stable keys; they are parameters, not operators.

Each element of the fragment contributes differently to the three RAI artifacts: the ASD, the ASRP, and the context capsule. 
In the ASD, \texttt{policy\_ok} is an admission precondition, while \texttt{admit}, \texttt{plan}, and \texttt{reserve} are permitted lifecycle actions and \texttt{commit} is the point where planning turns into actions that the service is accountable for. 
In the ASRP, the same elements become obligations that must be met whenever the corresponding step runs: an admission must be linked to the governing policy version established by the preceding policy check; a model-assisted plan must identify the model version and freshness of its inputs; a retryable reservation must carry an idempotency key or compensation obligation; and a commit must be uniquely associated with its authorization and selected plan.
In the context capsule, these obligations become concrete records. 
The emitted evidence set declares the minimum fields that the observer should find when the service follows this sequence at runtime.
A single ASC description therefore gives three related views of the same service: a descriptive one (the ASD), a normative assurance one (the ASRP), and a runtime evidence one (the context capsule).

ASC's type system is intended to reject violations such as crossing a domain without a mediated boundary, retrying a non-idempotent effect without a key, or recording a commit without its authorizing policy decision. These examples illustrate commit safety and policy continuity; additional checks cover lifecycle closure and resource-bounded evidence sufficiency.

\begin{figure*}[!t]
\centering
\includegraphics[width=0.86\textwidth]{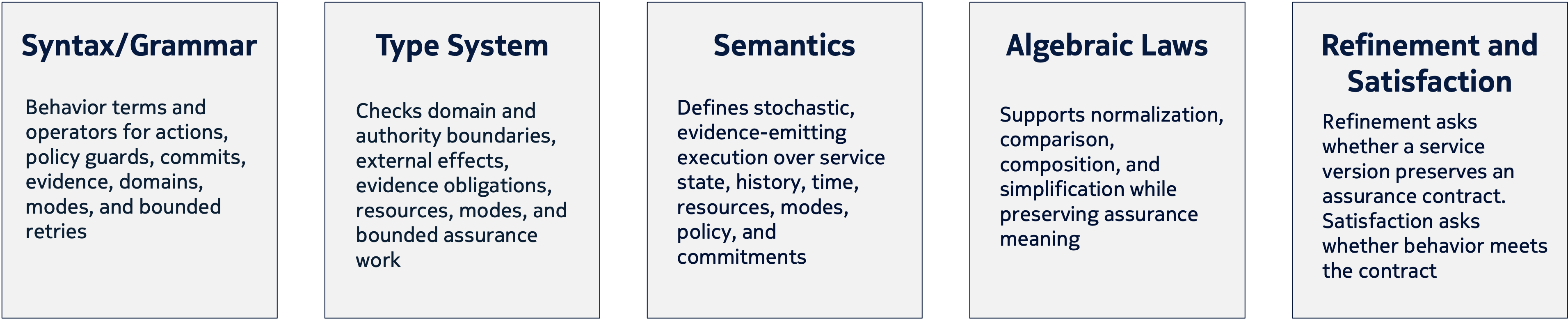}
\caption{Conceptual structure of the Agentic Service Calculus.}
\label{fig:asc-structure}
\end{figure*}

Fig.~\ref{fig:asc-structure} summarizes the five conceptual building blocks of ASC: syntax, a type system, execution semantics, algebraic laws, and refinement and satisfaction. The last item connects ASC to the artifact chain developed next: satisfaction concerns whether a service behavior meets its assurance contract, while refinement concerns whether a more specific or revised service description still preserves it.

\subsubsection{From ASD to ASRP}
RAI uses ASC to produce two artifacts. 
The \emph{Agentic Service Description} (ASD) is descriptive: it captures what a service may do, including its intents, actions, effects, modes, and policy boundaries.
The \emph{Agentic Service Reliability Profile} (ASRP) is the contract: it states what must be checked, recorded, recovered, and audited, which effects must be idempotent or compensable, and which degraded modes are admissible.

\begin{figure*}[!b]
\centering
\includegraphics[width=\textwidth]{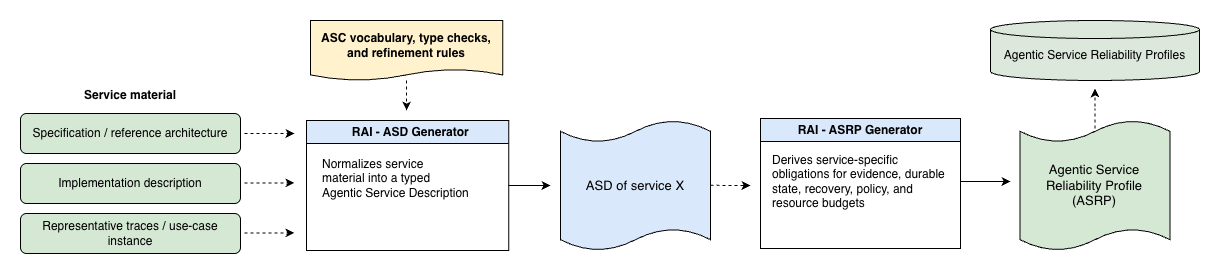}
\caption{Onboarding-time RAI pipeline: service material is normalized into an ASD and then compiled into a service-specific ASRP. Context capsules are produced later, at runtime, and do not appear here.}
\label{fig:rai-pipeline}
\end{figure*}

Separating description (ASD) from obligations (ASRP) supports versioning and composition: in the running example, the ASD captures the lifecycle sketched earlier, while the ASRP specifies which admission, reservation, commit, and model-dependent planning steps must remain durably accountable. 
A revised service can be checked against the prior ASRP to confirm it still meets the same obligations, and a cross-domain composition can rely on mediated steps and evidence envelopes without exposing each domain's internal trace. 
Fig.~\ref{fig:rai-pipeline} shows this onboarding-time pipeline, from the service's specification, implementation, and traces to the ASD and ASRP.

Together, these artifacts connect design-time modeling to runtime assurance: the context capsule, observer, and capsule controller then retain, verify, and adapt the evidence required by the ASRP.

\subsection{Assurance-Scoped Runtime}
\subsubsection{Context Capsules}
The \emph{context capsule} is the durable-state object specified by an ASRP. 
It preserves the subset of state needed to explain, replay, compensate, or audit the service rather than a complete transcript of every internal step. 
It is composed of layers that identify the record categories an ASRP may require. 
The intent layer records accepted scope and service identity. The plan layer records the selected plan, version, and input freshness. 
The commit and effect layers record commit boundaries, idempotency keys, confirmations, and compensation links. 
The policy and recovery layers record authority context, decisions, replay assumptions, and closure. 
The forensic layer is added only when risk or missing evidence warrants it.

A capsule policy is a service-specific contract, and how much it keeps scales with how much is at stake. 
A low-criticality service may retain only enough to identify the intent and its outcome, whereas a safety-sensitive service may also record the proofs, decisions, versions, and confirmations needed to justify and recover each action. 
Because the capsule can itself hold sensitive operational evidence, the ASRP also specifies its protection and retention requirements.

At runtime, the capsule adapts its recording mode to the current level of risk. 
Under normal conditions it uses a compact mode. 
When warning signs appear, such as repeated retries, stale information, or missing effects, it switches to an enriched or forensic mode for a bounded interval, then returns to compact mode once the situation stabilizes and the service can again be reliably reconstructed. 
The guiding idea is that recording should follow assurance needs, not merely the reflex to log more during incidents.

\subsubsection{Observer and Capsule Controller}
The {\em observer\/} checks the events and durable records that the service emits against the active ASRP and reports assurance findings, such as missing evidence, unauthorized transitions, stale observations, or retries that could repeat an external effect without a matching commit. It only judges conformance; it does not plan, select tools, or control the managed service. 
Whatever method it uses, its findings must stay auditable and within the ASRP.

The capsule controller adjusts how much the service records. 
Given the observer's findings, the current risk signal, and the durable-state budget, it selects the least costly recording mode that still satisfies the ASRP, which may mean keeping
the current mode unchanged. 
It records less when conditions are nominal and more when risk warrants it, from a compact capsule to enriched or forensic detail.

The control invariant is that every recording mode preserves the evidence the active ASRP makes mandatory. 
Budget pressure may drop optional diagnostic or forensic detail, but never required authorization, commit, effect, compensation, or disposition records.
Escalation and de-escalation use separate thresholds, hysteresis, and minimum dwell times, escalating when risk or evidence deficiency rises and de-escalating only after risk decays and replay sufficiency is restored. 
This keeps the controller's work bounded and prevents oscillation.

\begin{figure*}[!b]
\centering
\includegraphics[width=0.88\textwidth]{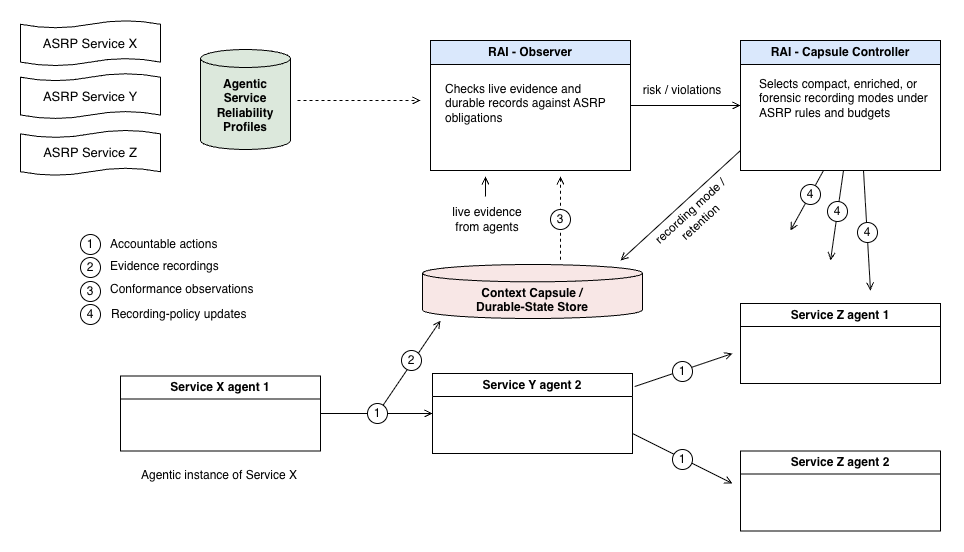}
\caption{Evidence-centered reliability loop: the observer checks evidence against the ASRP, and the capsule controller adjusts recording within a bounded budget.}
\label{fig:runtime-loop}
\end{figure*}

The loop makes both service performance and evidence sufficiency controlled variables. 
The question is not only whether a service met its latency or availability objective, but also whether enough trustworthy state remains to recover, audit, and justify the behavior under the active mode and budget. 
Fig.~\ref{fig:runtime-loop} summarizes this loop.

\section{Operational Integration}
RAI can be introduced without replacing telemetry, service-assurance platforms, or management APIs. 
The natural integration point is the durable-state boundary where an agentic management system already records intents, reservations, actions, and outcomes. 
The service emits the evidence required by its ASRP, and the observer checks conformance. 
High-criticality actions may require a synchronous check before commit; lower-criticality actions can use asynchronous checking so the observer stays off the critical path.

At onboarding time, RAI acts as an assurance compiler. 
A service specification, reference architecture, or implementation description is translated into an ASD and checked for domains, authority, external effects, resource limits, evidence obligations, and degraded modes.
The resulting ASRP can generate capsule schemas, logging adapters, retention rules, runtime assertions, and deployment gates. 
The strongest such gate is simple: a system that cannot exhibit a valid profile for its external effects and degraded behavior should not be promoted to unrestricted autonomous operation.

Deployment can be staged: a first version checks only that every accepted intent has an admission record, plan identifier, commit key when an external effect occurs, and final disposition; later versions add policy-context and stale-observation checks, bounded retries and compensation links, and finally adaptive capsule modes. 
This strengthens existing observability with assurance semantics rather than imposing a new monolithic platform.

Cross-domain operation relies on mediated evidence. A peer domain need not expose its internal topology, models, or policies; instead, it returns a compact evidence envelope that reports only what assurance needs, such as the commitments it made and the effects it confirmed. 
A boundary mediator handles the incoming intent, applies local policy, and returns an assurance record for the upstream capsule. 
Responsibility is shared, so each team owns the artifact that pertains to its role: the ASD for those who build the service, the ASRP for those who assure it, the durable store for those who operate it, and the policy and protection rules for those who secure it.

\section{Applying RAI to the Lifecycle Manager}
Table~\ref{tab:lifecycle-application} maps the running example's lifecycle to minimum durable records and representative observer checks. 
The table makes the example operational: the ASRP is not a generic instruction to ``log more,'' but a phase-specific contract.

\begin{table*}[!t]
\caption{Illustrative ASRP for the Agentic Lifecycle Manager}
\label{tab:lifecycle-application}
\centering
\footnotesize
\renewcommand{\arraystretch}{1.03}
\begin{tabularx}{\textwidth}{p{0.105\textwidth}YY}
\toprule
Phase & Minimum durable records & Representative observer check \\
\midrule
Admit & Accepted intent, admitted scope, policy version, admission proof. & Intent was admitted under an active or still-valid cached policy. \\
Plan & Plan identifier, observation freshness, model or planner version, safety-filter result when applicable. & Selected plan used admissible and sufficiently fresh inputs. \\
Reserve & Tentative reservation, peer-domain evidence, idempotency or compensation link. & Reservation is confirmed, safely repeatable, or reversibly tentative. \\
Commit & Durable commit key bound to the selected plan and authorization. & A unique accountable execution boundary exists before external effects. \\
Execute & Effect record, idempotency key, outcome confirmation. & Duplicate effects are prevented or detected and compensated. \\
Degrade & Trigger, restricted contract, cached-policy validity, authority and recovery conditions. & Degradation does not weaken safety, authority, or evidence obligations. \\
Audit & Final disposition, open-obligation closure, retention decision. & The intent lifecycle is closed or explicitly escalated. \\
\bottomrule
\end{tabularx}
\end{table*}

Suppose a high-load interval produces repeated planning retries and packet latency approaches its bound. 
Conventional monitoring reports queueing and planning-retry counts. 
Beyond these metrics, the observer can detect assurance-relevant problems that they miss, such as retries never matched by a commit or an admission that degraded without a recovery path. 
In response, the capsule controller temporarily enriches recording for the affected service alone, capturing the extra evidence needed to explain and recover the episode, returning to compact mode once conditions recover.

The example also shows why full-trace retention should not be the default. 
RAI keeps the records required to support an assurance claim rather than every intermediate step. 
Most candidate plans and the planner's internal deliberation can be omitted. 
For a selected plan, the model version, relevant inputs, safety result, and commit boundary may provide sufficient evidence for that claim.

\section{Validation and Open Challenges}
This section outlines a validation path for RAI and highlights the main open challenges.

\subsection{Validation Methodology}
This article presents RAI as an architectural proposal. 
Its validation should therefore make clear what would confirm or refute the claims.

\emph{Structural validation} tests whether a service's specification and implementation can be turned into an ASD and ASRP, and whether RAI's checks surface the parts of the service's behavior that matter for assurance, such as commit points, authority boundaries, and degraded modes. 
\emph{Runtime validation} replays recorded traces through the observer and measures how reliably it detects assurance failures, such as missing evidence or a duplicated external effect. 
\emph{Resource validation} weighs the savings from selective recording against the assurance it preserves, for example the capsule's size relative to a full trace and whether incidents can still be replayed. 
\emph{Governance validation} injects policy brownouts and cross-domain conflicts and checks that restricted modes still preserve authority, auditability, and required evidence.

Digital twins and controlled sandboxes provide suitable environments for validating RAI under controlled failure injection, because coupled failures are rare and difficult to reproduce from operational traces. 
A first validation step is to compare three recording strategies: (i) the full trace, which explains everything but is costly; (ii) a fixed minimal record, which is cheap but can miss rare obligations; and (iii) the adaptive capsule governed by the ASRP, which aims for the explanatory value of the full trace at closer to the cost of the minimal one. 
Each strategy would be exercised under both normal operation and stress, such as a retry storm or a policy brownout.

This validation step tests five explicit hypotheses. 
\emph{H1: storage efficiency.} Relative to full-trace recording, the adaptive capsule reduces bytes written per completed intent and total retention cost.
\emph{H2: assurance preservation.} Relative to a fixed compact policy, under elevated-risk workloads the adaptive capsule retains more of the evidence that assurance requires and enough to reconstruct what happened. 
\emph{H3: assurance-specific detection.} The observer detects missing commits, stale decision context, and policy ambiguity that ordinary performance telemetry does not identify. 
\emph{H4: controller stability.} Adaptive recording remains stable, as measured by mode transitions per interval, mode dwell time, and the rate of false escalation. 
\emph{H5: recoverability.} Reusing the same keys across retries and commits makes it easier to detect when an external effect was applied more than once, and to recover from it, either by undoing the extra effect (compensation) or by re-running the sequence to a correct state (replay).
Each hypothesis is evaluated under the same intent workload and failure injection so that differences can be attributed to the recording policy rather than to different executions.

The hypotheses also define failure conditions: H1 fails if the adaptive capsule nears full-trace volume in ordinary operation or does not cut retention cost; H2 if savings come from losing mandatory evidence or an incident cannot be replayed; H3 if observer findings add nothing beyond conventional telemetry; H4 if the controller oscillates or stays
elevated after risk decays; and H5 if duplicate effects or the required recovery paths cannot be reconstructed. 
A restricted mode also fails if it remains available only by dropping required policy, authority, commit, or audit evidence. 
The aim is the best recoverability and accountability within the durable-state budget.

\subsection{Open Challenges}

RAI can advance assurance-scoped reliability in agentic networks, but several challenges still require further study.

The first challenge is formal tractability. 
ASC must expose enough structure for typing, refinement, composition, and resource reasoning while staying practical, so that onboarding a service does not require constructing, by hand, a formal proof that its description satisfies these assurance properties. 
The immediate next steps are to find restricted forms of ASC, using only a subset of the language's constructs, that can still describe realistic services but are simple enough for a tool to check automatically, and to build those tools.

The second challenge is cross-domain semantics. 
End-to-end assurance must be constructed from evidence envelopes and mediated commitments when direct visibility is impossible. 
The open question is how much semantics must cross a boundary to support an assurance claim without exposing unnecessary operational or business detail.

The third challenge is stability and self-assurance. 
Adaptive recording must not amplify load when the managed network is already stressed. 
A cap on the extra work that recording can add, hysteresis to avoid flipping between modes, and per-service priorities are required to prevent oscillation and runaway recording. Under overload or loss of required assurance evidence, the capsule controller should enter a predefined restricted mode. 
In that mode, it preserves mandatory assurance records, suspends actions whose assurance conditions cannot be established, and escalates when necessary. 
This adapts the fail-safe design principle of IEEE Std~7009~\cite{ieee_7009_2024} to assurance control.

The fourth challenge combines evidence minimization, privacy, and emergent behavior. 
Capsules must retain enough to reconstruct an assurance account without themselves becoming ever-growing stores of sensitive operational data. 
Multi-agent compositions can also fail through oscillation, correlated retries, or delayed feedback even when each component locally conforms. 
These issues intersect with the governance and risk-management practices of the NIST AI RMF~\cite{nist_ai_rmf_2023}.

\section{Conclusion}
Agentic network and service management makes reliability depend on whether autonomous behavior can be justified by sufficient evidence and constrained by policy. 
Uptime and telemetry remain necessary, but they show only that the system is running, not that its autonomous decisions were properly authorized, accountable for their effects, and backed by enough evidence to recover and audit them.
RAI provides a general artifact chain for these questions. 
ASC supplies a common assurance language, the ASD states what a service may do, and the ASRP states what must be assured. 
The context capsule and the evidence-centered feedback loop between observer and capsule controller then apply those obligations to runtime durable-state control. 
The deterministic-service lifecycle manager example illustrates how profile-selected evidence can be richer than a conventional log while avoiding indiscriminate full-trace retention. 
As agentic systems move from scripted workflows toward goal-oriented operation, recording the state that matters, and being explicit about why, can become a practical foundation for trustworthy autonomy.

\section*{Acknowledgments}
The authors acknowledge Gary Atkinson for leading the Reliable Agentic Networking project at Nokia Bell Labs, within which this work was developed.

\section*{Biographies}
{\footnotesize
\setlength{\parindent}{0pt}
\setlength{\parskip}{0.65\baselineskip}

\textbf{Bilgehan Erman} (bilgehan@ieee.org) is a Senior Researcher at Nokia Bell Labs. His recent work focuses on autonomous and agentic network systems, with emphasis on formal assurance frameworks, reliable AI-driven operations, and trustworthy network automation. He holds an M.S. in Computer Engineering and a B.S. in Electrical Engineering, both from METU in Ankara, Türkiye.

\textbf{Andrea Francini} (andrea.francini@nokia-bell-labs.com) is a Distinguished Member of Technical Staff (DMTS) at Nokia Bell Labs, where he leads activities on deterministic networking, network-application symbiosis, and congestion control for AI data centers. He holds a Ph.D. in Electrical Engineering from the Politecnico di Torino, Italy.

\textbf{Nikos Papadis} (nikos.papadis@nokia-bell-labs.com) is a Research Scientist - Networks Researcher at Nokia Bell Labs in NJ, USA. He received the M.Sc., M.Phil., and Ph.D. degrees in Electrical Engineering from Yale University, CT, USA, and his Diploma in Electrical and Computer Engineering from the National Technical University of Athens, Greece. His research interests span next-generation and agentic networks, federated learning and machine learning more broadly, blockchain technology, and payment networks.
\par
}
\end{document}